\documentclass[aps,floats,showpacs,noshowkeyws,preprintnumbers,nofootinbib]{revtex4}

\usepackage{bbm}                                                  
\usepackage{color}
\usepackage{graphicx}
\usepackage{bm}

\newcommand{\nc}{\newcommand}
\nc{\beq}{\begin{equation}}  \nc{\eeq}{\end{equation}}
\nc{\bea}{\begin{eqnarray}}  \nc{\eea}{\end{eqnarray}}
\nc{\bit}{\begin{itemize}}  \nc{\eit}{\end{itemize}}

\def\mBB{{\mathbbm M}}
\def\bBB{{\mathbbm B}}
\def\sm{Standard Model}
\def\lcal{{\cal L}}

\def\ncal{{\cal N}}
\def\half{\frac12}
\def\phibf{{\bm\phi}}
\def\nn{{\bf n}}
\def\deriva#1#2#3{\left(\frac{\partial #1}{\partial #2}\right)_{#3}}
\def\inv#1{\frac1{#1}}
\def\aa{{\bf a}}

\def\sqr#1#2{{\vcenter{\hrule height.#2pt
      \hbox{\vrule width.#2pt height#1pt \kern#1pt
         \vrule width.#2pt}
      \hrule height.#2pt}}}
\def\square{\mathchoice\sqr56\sqr56\sqr{2.1}3\sqr{1.5}3}

\def\Phibf{{\bm\Phi}}
\def\FF{{\bf F}}
\def\then{{\quad\Rightarrow\quad}}
\def\ff{{\bf f}}
\def\chibf{{\bm\chi}}
\def\vevof#1{\left\langle #1 \right\rangle}
\def\cc{{\bf c}}
\def\up#1{^{\left( #1 \right) }}
\def\vv{{\bf v}}

\def\mati{{\mathbbm1}}
\def\BB{{\bf B}}
\def\vcal{{\cal V}}
\def\bcal{{\cal B}}
\def\XX{{\bf X}}
\def\ss{{\bf s}}

\def\phit{{\tilde\phi}}
\def\phitbf{{\tilde{\phibf}}}
\def\curlh{{\tilde\gamma}}
\def\dh{k}
\def\dH{K}
\def\ds{\tau}

\def\mpl{M_{\rm Pl}}
\def\tP{\tilde\Phi}
\def\tPbf{\tilde\Phibf}

\def\Or{{\rm O}}

\def\xu{\vartheta}

\begin{document}

\title[Periodic thick-brane configurations]{Periodic thick-brane configurations and their stability}

\author{Jos\'e Wudka}
\email{jose.wudka@ucr.edu}
\affiliation{Department of Physics, University of California,
Riverside CA 92521-0413, USA}

\begin{abstract}
I investigate models with scalar fields
in 5 dimensions that exhibit thick-brane 
configurations with a non-trivial metric. 
I show that an
appropriate coupling to the scalar curvature allows for
periodic configurations, which, however, are
unstable under small harmonic perturbations.
A model for stabilizing these
configurations is proposed and discussed.
\end{abstract}


\maketitle

\bigskip

\section{Introduction}
\label{sec:intro}

One of the most interesting proposed  extensions of the
standard model is based on a 5-dimensional space-time
\cite{rs} containing one or more
branes, 4-dimensional subspaces where some of the fields are
localized. Such models usually assume that space time has a topology $
\mBB \times \bBB$, where $ \mBB$ denotes the usual Minkowski space and $\bBB$
is a manifold that may or may not be compact, and a non-separable
metric (I consider here only the so-called ``small extra
dimension'' models); the original models assumed $\bBB$ to be one-dimensional
but this has since been generalized \cite{d.6}.
When this space-time configuration and brane content correspond to a
stable solution of the field equations, it can serve as a vacuum
where quantum fluctuations propagate; the assumption being that this
background configuration corresponds to a minimum of the effective
action for the complete system.

This paradigm offers an innovative solution to the hierarchy
problem and a wealth of new effects that may be observed
 at the LHC.  Motivated by this, several authors have
constructed realistic or semi-realistic models based on this idea,
and have studied a variety experimental signatures
\cite{phen}. 

The stability of the background
configurations under small harmonic perturbations 
has also been studied \cite{rs,rad.stab,rad}. In the seminal 
publications~\cite{rs} it was shown that for the models there considered
most perturbations are not destabilizing; the one
exception being the dilaton, a self-similar scaling perturbation
which is neutral (has no quadratic potential term or,
equivalently, is associated with zero frequency). This problem, however
can be eliminated by the addition of appropriate
bulk scalar fields \cite{gw}, which has been studied in a variety of
cases: for single scalar fields coupled minimally to gravity
\cite{gw,BSMP}, non-minimally coupled scalars \cite{BSNP}, and for
Brans-Dicke theories \cite{BSBP}.

These higher-dimensional 
theories, being non-renormalizable, have an intrinsic ultraviolet cutoff
scale $ \Lambda $ beyond which they are not reliable (at least within perturbation
theory).
In models where branes are infinitely thin (in the fifth coordinate),
it is tacitly assumed that their structure will become manifest at
scales $ \gtrsim \Lambda $. But this need not be the case, and a
certain amount of attention has been paid to the possibility that the
dynamics at scales below $ \Lambda $ is responsible for the brane
 configuration.  A variety of such models have been studied in the
literature for non-compact $\bBB$ with one \cite{NSMNN,NSMNS,pot}, or
several \cite{NMMNN,NMMNS} minimally-coupled scalars.  In addition,
the stability
of such models has been investigated for the case of non-compact $\bBB$ and
one \cite{NSMNS,pot} or several \cite{NMMNS} minimally coupled
scalars, and for $\bBB$ compact and a flat metric \cite{bg}.

Models
with $\bBB$ compact that exhibit periodic, stable,
brane-like configurations 
generated by physics below $ \Lambda $
(``thick branes''), and which have a non-trivial
metric are more difficult to construct. The reasons are,
first, that for the simplest models the background configurations
cannot satisfy the periodicity requirement ({\it e.g.}
they fail to satisfy one or more of the sum rules 
listed in \cite{srules}, as required for consistency). 
And second, general considerations
\cite{bg,persol} apparently preclude the stabilization of
thick-brane solutions using only scalar fields.

The goal of the present paper is to provide
mechanisms that overcome these two difficulties. 
Specifically, to exhibit a class of 5-dimensional models
containing gravity, scalars, and 
antisymmetric tensor fields, which
admit periodic kink-like solutions that are perturbatively
stable at scales below $ \Lambda$. Though the theories
thus obtained have no phenomenological applications 
({\it e.g.} they exhibit but a small amount of warping,
so that no significant mass hierarchy can be generated),
they are of interest 
because they can address these two problems. Though
there are some indications
that more realistic configurations can also be obtained
(sect. \ref{sec:bs}),
I will not attempt to construct a phenomenologically viable model,
nor will I not attempt to address the much more ambitious and difficult
problems of confining fields to the branes and of global stability. 
For a different
class of theories, based on the ``large extra dimensions''
paradigm~\cite{add}, there are realistic models that include both dynamically
generated branes and a confining mechanism for the \sm\ fields; see
for example~\cite{rv}.

The calculations presented are essentially
classical, amounting to obtaining solutions to the
equations of motion that are stable under harmonic
perturbations. The usefulness of these results
lie in the well-established connection between such solutions
and related quantum objects~\cite{Jackiw:1977yn}; in particular,
the classical background solutions are to be interpreted
as the lowest order semi-classical approximation 
to a quantum vacuum, and the
frequencies of the harmonic perturbations as the energies of
the low-lying excitations.

The plan of this paper is the following. The next section introduces a 
class of models involving only scalars, 
that support periodic configurations, which in
favorable cases are similar to the ones in \cite{rs}. These
background configurations are unstable
(Sect. \ref{sec:stability}), but this problem can be solved
through the introduction of antisymmetric
tensor fields (Sect. \ref{sec:atf.stab}), adequately coupled. The last section
contains some parting comments and observations, while some
mathematical considerations are delegated to the appendices.

\section{The model (basic version)}

I first consider the problem of constructing kink-like configurations
that involve gravity and are periodic in the fifth
coordinate of a 5-dimensional space-time\footnote{In the following
$i,j,$ etc denote 5-dimensional space-time indices, Greek indices will
refer to the 4-dimensional non-compact coordinates; the metric
signature is $ (-1, +1, +1, +1, +1) $. I use the conventions of
Landau and Lifshitz \cite{ll} for the definition of the Riemann and
associated tensors. I denote the compact coordinate by $y$, and the
non-compact ones by $ x^\mu, ~ \mu=0,1,2,3 $. The indices $r,s,$
etc. label the scalar-field components.}.
Though it is well known
\cite{sol} that scalar models with Lagrangians of the type $ \lcal
\sim ( \partial\phi)^2 - V( \phi ) $ do allow stable kink and
multi-kink configurations, none of these satisfy the periodicity
constraint~(the simplest way of seeing this is by noting that
such configurations violate one or more of the sum rules of Ref
\cite{srules}, as shown in that same publication); fortunately  a
simple and natural modification overcomes this obstacle.  I consider
models of the type~\cite{BSNP},
\beq
\lcal = 2\mpl^3 R - \half g^{ij} \phit_{r,i} \phit_{r,j} - \half
\xi_{r s}( \phit_r + \hat n_r \varphi)
         ( \phit_s + \hat n_s \varphi) R 
+ V \left( \phitbf \right)
\,,
\label{eq:l.phi}
\eeq
where $\mpl$ denotes the 5-dimensional Planck mass, and 
$R$ the scalar curvature,
a comma denotes an ordinary derivative,   
$ \phitbf $ is an $N$ component real scalar field 
with components $ \phit_r $,
and $\xi$ is a real, symmetric $ N \times N $ matrix.  
The vector $ \hat\nn$
denotes a constant direction in field space, while 
$ \varphi $ is a scale of the same order as 
$ \mpl^{3/2}$. The sign of the potential term is
chosen for later convenience.

Models of the type (\ref{eq:l.phi}) have been
studied in the literature, in particular Ref.~\cite{bg} 
provides flat-space, single-field examples 
where the background solutions are both periodic and stable;
unfortunately these models do not extend easily to the
case of a non-trivial metric.

I assume now that $ \mpl $ is the largest scale in the theory
and consider solutions that  
have a small but non-trivial deviation from a flat metric
(recalling that $ \varphi \sim \mpl^{3/2}$):
\bea
\phitbf &=& \phibf + O(1/\varphi)\,; \cr
g_{ij} &=& \eta_{ij} + \inv\varphi h_{ij} + 
\inv{\varphi^2} \dh_{ij} + O(1/\varphi^3) \,;
\label{eq:expansion2}
\eea
where $ \eta_{ij} = {\rm diag}(-1,1,1,1,1) $ denotes
the flat-space metric.
Substituting in (\ref{eq:l.phi}) gives,
after some algebra,
\bea
\sqrt{-g} \, \lcal &=& - \half \sum m_{rs} \phi_{r,i} \phi_r{}^{,i} 
+ V(\phibf) \cr && \quad
- \frac8{3\bar\xi}
\left[ \bar\xi \hat\aa \cdot \phibf_{,i} -
\frac3{16}(\aa\cdot\hat\nn) q_i \right]
\left[ \bar\xi \hat\aa \cdot \phibf^{,i} -
\frac3{16}(\aa\cdot\hat\nn) q^i \right] \cr
&& \quad + \frac{|\aa|^2}{8\bar\xi}\left( h^{ij,k} h_{ij,k}
- h^{ik}{}_{,k}h_{ij}{}^{,j} - \inv4 q_i q^i \right) + O(1/\varphi) \,,
\label{eq:l.pert}
\eea
where space-time
indices are raised and lowered using the flat metric $ \eta $,
and where
\bea
q_i = h^k_{i,k} - h^k_{k,i} \,, \quad && \quad
{\rm a}_r = \xi_{ r s} \hat n_s \,, \cr
m_{rs} = \delta_{rs} - \frac{16}3 \bar\xi \hat{\rm a}_r \hat{\rm a}_s \,, 
\quad && \quad 
\bar\xi = \frac{|\aa|^2}{\aa \cdot \hat\nn - 4\mpl^3/\varphi^2} \,.
\label{eq:q.m}
\eea

The corresponding equations of motion are:
\bea
\qquad\quad m_{rs} \square\, \phi_s  + \deriva V{\phi_r}{} &=&0\,; \qquad
\square = \eta^{ij} \partial_i \partial_j \,,\cr
\square\, h_{ij} - h^k_{i,j k} - h^k_{j,i k}
+ h^k_{k, i j} &=& - 4 \frac{\bar\xi}{| \aa|}
\left[ \left( \hat\aa \cdot\phibf \right)_{, ij}
 + \inv3 \eta_{ij} \square \left( \hat\aa \cdot\phibf \right)
\right] \,;
\label{eq:pert.eom}
\eea
which can also be obtained  by expanding the Einstein and
field equations and using (\ref{eq:expansion2}).

\subsection{Background solutions}
\label{sec:bs}

With the goal of preserving 4-dimensional Lorentz invariance I
look for background configurations of the form
\bea
\phibf &=& \Phibf(y) \,, \cr 
h_{ij} &=& H_{ij}(y), \quad
H_{\mu\nu} =  -2 \sigma(y) \eta_{\mu\nu}, 
\quad H_{4 \mu} = H_{44} =0 \,; \cr
\dh_{ij} &=& \dH_{ij}(y), \quad
\dH_{\mu \nu} =  -2 \ds(y) \eta_{\mu\nu}, 
\quad \dH_{4 \mu} = \dH_{44} =0 \,;
\label{eq:back}
\eea
which  solve  (\ref{eq:pert.eom}) to lowest order in $ \varphi$ provided
\bea
 \sigma= \frac23 \frac{\hat\aa \cdot \Phibf }{\hat \aa \cdot
\hat\nn} + \hbox{const.}\,, \quad && \quad
\Phibf' m \Phibf' 
=
\left[ -\frac{27}2(\aa\cdot\hat\nn)
\sigma^2
 + \Phibf \xi \Phibf - 3 (\aa\cdot\hat\nn) \ds \right]'' \,, \cr
 && \cr
 \half \Phibf' m \Phibf' + V \left( \Phibf\right)
=0 \,, \quad &&\quad 
 m \Phibf'' = \FF; \quad F_r = - \deriva
V{\phi_r}{\hbox{\scriptsize$\phibf$} = \Phibf} \,;
\label{eq:back.0}
\eea
where a prime denotes a $y$ derivative.

The last two equations in (\ref{eq:back.0}) describe the zero-energy,
classical, non-relativistic motion in a potential $V$; the mass
(cf. eq. \ref{eq:q.m}) equals one except when $
\Phibf $ is parallel to \aa, in which case the mass is $1 -
(16/3)\bar\xi $ 
(for a different interpretation when $ \bar\xi  > 3/16 $,
see  \ref{sec:app.c}). Higher-order
corrections (in $1/\varphi$)
can be determined similarly, but the expressions are
cumbersome; an example is presented in \ref{sec:app.a}.
 
The warp factor in this type of models is
$ \sim 	1 -  2 \sigma/\varphi$, and cannot
generate a significant mass hierarchy. Note
however that the expression in \ref{sec:app.a} for $ \tau $
shows that this function is also periodic, with the 
same period as $ \sigma $; which suggests that
periodic solutions
exist also for moderate values of $ \varphi $. Verification
of this conjecture, as well as a determination of the 
range in $ \varphi $ for which it holds, lies beyond
the scope of this paper.

 In the following I will choose coordinates in field space such that
\beq
\hat{\rm a }_r = \delta_{r,1}
\then
m = {\rm diag}\left(1 - \frac{16}3 \bar\xi,1,1, 1,\cdots \right) \,,
\label{eq:m.diag}
\eeq
and will denote by $ m_r$ the eigenvalues of $m$:
\beq
m_1 = 1 - \frac{16}3\bar\xi \,, \qquad m_r=1,~~ 2\le r \le N
\eeq

A judicious choice of the potential $V$ 
and initial conditions will lead to solutions that are
periodic in $y$,
\beq
\Phibf(y+L ) = \Phibf(y) \,,
\label{eq:per.con}
\eeq
with $L$ determined by $V$ and the initial conditions;
$ \sigma $ will also be periodic with the same period.
For example, adopting the basis (\ref{eq:m.diag}) and taking
\beq
V(\phibf) = \sum_r m_r u_r \left( \phi_r \right) \,,
\label{eq:simple.v}
\eeq
the equations (\ref{eq:back.0}) yield
\beq
\sigma = 2 \Phi_1/(3\hat n_1) \,;\qquad
\Phi_r{}'' =  f_r \left(\Phi_r\right)\,;
\label{eq:simple.back}
\eeq
where $ f_r = - d u_r(\phi_r)/ d \phi_r $ (no sum over $r$). An
appropriate choice of $ u_r $ and initial conditions will then
generate solutions obeying (\ref{eq:per.con}).

Adopting such solutions as background configurations allows the
identification $ y \equiv y~{\rm mod}~L $, which amounts to
a compactification of the fifth dimension. Such
configurations can serve
as vacua for the theory, provided they are stable.

\subsection{Sum rules}
\label{sec:sr}

It is of interest to see how the present models avoid the obstacles
listed in Ref. \cite{srules}. The most severe constraint is obtained by
assuming the existence of periodic solutions and integrating the
second equation in (\ref{eq:back.0}) over a period:
\beq
\oint dy \, \Phibf' m \Phibf' = 0  \,.
\label{eq:sum.rule}
\eeq
When $ \bar\xi =0 $ this can be
satisfied only
 when $ \Phibf' =0 $ (in the presence of localized branes
there are additional contributions and the sum-rule can
be satisfied by non-trivial $ \Phi $ configurations~\cite{srules}). 
In contrast, when $ \bar\xi\not=0 $,
 non-trivial solutions are allowed provided $m$ is not
positive definite; from (\ref{eq:m.diag}) this corresponds to
\beq
\bar\xi > \frac3{16} \,,
\label{eq:xi.range}
\eeq
which I assume henceforth. In this case the mass matrix
(\ref{eq:m.diag}) takes the form
\beq
m = |m| S; \quad  S = {\rm diag}(-1,+1,+1, \cdots) \,.
\label{eq:def.of.S}
\eeq

\subsection{Comparison with the 2-brane Randall-Sundrum model}
\label{sec:rs}
In the 2-brane Randall-Sundrum model \cite{rs} the background of the form
(\ref{eq:back}) is also adopted, but instead of introducing scalar fields it
is assumed that the space-time contains two branes with cosmological
constants $ \pm \lambda $, and a bulk cosmological constant $ \Lambda
< 0 $. The resulting field equations are
\beq
24 \mpl^3 \sigma'{}^2 = - \Lambda \varphi^2 \,, \qquad
12 \mpl^3 \sigma'' = \lambda \varphi \left[ \delta(y) -
\delta(y-L/2)\right] \,,
\label{eq:rs}
\eeq
where $0\le y \le L$ is the range of the compact coordinate.

\begin{figure}[ht]

\begin{center}
\includegraphics[width = 3in]{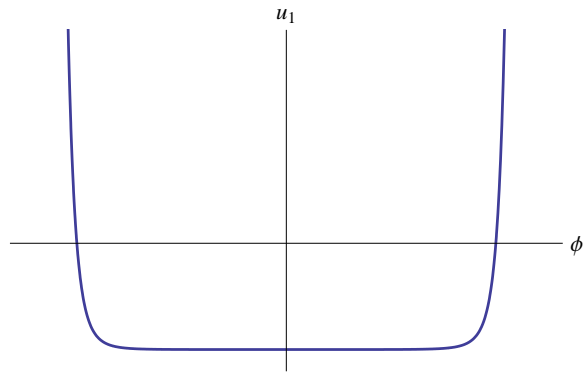}
\end{center}

\caption{Scalar potential leading to configurations similar to those
described in \cite{rs}}
\label{fig:i}
\end{figure}

In order to compare this to the previous results I again adopt 
(\ref{eq:m.diag}) and assume the
potential takes the form (\ref{eq:simple.v}). Then, from
(\ref{eq:back.0}),
\beq
24 \mpl^3 \sigma'{}^2 = \frac{32 \mpl^3}{3 \hat n_1^2} \Phi_1' {}^2 \,, \qquad
12 \mpl^3 \sigma'' = \frac{8\mpl^3}{ \hat n_1}\Phi_1'' \,.
\label{eq:mock.rs}
\eeq
Using (\ref{eq:simple.back}), it is clear that for a very flat
potential $ u_1 $ (see Fig. \ref{fig:i}), $ \Phi_1'$ will be
almost constant except at the turning points where it will rapidly
drop to zero, while $\Phi_1''$ will be almost zero except at the
turning points, where it will be large. This type of potential then
yields configurations qualitatively similar to those derived from
(\ref{eq:rs}) for the case $ \Lambda < 0 $.

\section{Stability}
\label{sec:stability}

The usefulness of the above solutions as background configurations 
depends on their stability; the minimal requirement being that 
all periodic solutions to
the linearized perturbation equations are bounded in
time, {\it up to coordinate transformations}.  
As usual, such linear perturbations can be assumed to depend harmonically
on the non-compact coordinates.

The relevance of coordinate transformations can be illustrated by
considering a background solution of period $L$,
\beq
\Phibf (y) = \sum \ff_n e^{ 2 \pi i n y/L} \,;
\qquad
\sigma (y) = \sum \ss_n e^{ 2 \pi i n y/L} \,;
\eeq
and a perturbation that consists in
the replacement $ L \to (1-\epsilon)L $:
\bea
\delta_L \phibf &=& \left[ \Phibf \right]_{L \to L - \epsilon L} - \Phibf =
\epsilon y \Phibf'  - \epsilon L \sum e^{ 2 \pi i n y/L} \partial_L \ff_n \,, \cr
\delta_L g_{\mu \nu} &=& - 2 \epsilon y \sigma' \eta_{\mu \nu} 
 - 2\epsilon L \sum e^{ 2 \pi i n y/L} \partial_L \ss_n  \eta_{\mu \nu} \,,
\cr
\delta_L g_{4 \mu} &=& \delta_L g_{44} =0 \,.
\label{eq:period.pert}
\eea
Though this suggests the need to include periodic
perturbations multiplied by linear functions of $y$, this is not the
case since these modes can be absorbed using appropriate coordinate
transformations. Under $ x^i \to x^i + \xi^i(x) $,
\bea
\delta_{\rm coord} g_{ij} &=& \xi_{i;\,j} + \xi_{j;\,i} + O ( \xi^2) \,, \cr
\delta_{\rm coord} \phibf &=& \xi^i \partial_i \Phibf + O ( \xi^2) \,,
\label{eq:coord.transf}
\eea
where a semicolon denotes a covariant derivative, 
and choosing $ \xi_4 = - \epsilon y
,~ \xi_\mu =0 $, a combination of (\ref{eq:period.pert}) and
(\ref{eq:coord.transf}) gives
\bea
\delta_L' g_{44} &=& - \epsilon \,; \qquad
\delta_L' g_{ 4\mu} = 0 \,; \cr
\delta_L' g_{\mu \nu} &=&  
 - 2\epsilon L \sum e^{ 2 \pi i n y/L} \partial_L \ss_n  \eta_{\mu \nu} \,, \cr
\delta_L'\phibf &=& - \epsilon L \sum e^{ 2 \pi i n y/L} \partial_L \ff_n \,,
\label{eq:delta.prime}
\eea
where $ \delta_L' = \delta_L + \delta_{\rm coord}$. The variations $ \delta_L'
g_{ij},~ \delta_L'\phibf $ are then equivalent to
(\ref{eq:period.pert}) and are periodic in $y$.

Hence I will 
look for solutions to  (\ref{eq:pert.eom}) of the form 
\bea
\phibf &=& \Phibf(y) + e^{i p \cdot x} |m|^{-1/2} \chibf(y)\,,
\quad  p \cdot x = p_\mu x^\mu \,; \cr
h_{ij} &=& H_{ij}(y) + e^{ i p\cdot x } \gamma_{ij} (y)  \,,
\label{eq:pert.expansion}
\eea
keeping only first-order terms in the 
perturbations $ \chibf,~\gamma_{ij} $ (the matrix
$|m|^{-1/2} $ -- cf. eq. \ref{eq:def.of.S} -- is included for
later convenience).

In order to
simplify the calculations it is convenient to choose coordinates
such that
\beq
\gamma_{4 \mu } = 0 ; \quad \gamma_i^i = 0\,.
\label{eq:g.conditions}
\eeq
(as before, indices are raised and lowered using the flat-space metric).
These conditions, however, do not completely fix the coordinates: 
using (\ref{eq:coord.transf}) with
\beq 
\xi^4 = i X'(y) e^{ i p \cdot x}, \quad
\xi^\mu = X(y) p^\mu e^{ i p \cdot x}  \,; \quad 
X'' + p^2 X =0, ~~X \sim1/\varphi \,,
\label{eq:new.x1}
\eeq
yields $\gamma_{\mu \nu} \to \gamma_{\mu\nu} + 2i p_\mu p_\nu X$ and
$ \gamma_{44} \to \gamma_{44} + 2 i X'' $, which
preserve (\ref{eq:g.conditions}).

Expanding 
\beq
\gamma_{\mu\nu} = \frac{p_\mu p_\nu}{p^2} \Gamma_L 
+ \left( \eta_{\mu\nu} - \frac{p_\mu p_\nu}{p^2} \right)  \Gamma_T 
+ \curlh_{\mu\nu}; \quad p^\mu p^\nu \curlh_{\mu\nu} =
\eta^{\mu \nu} \curlh_{\mu \nu} = 0  \,,
\eeq
where $ p^2 = 
\eta^{\mu\nu} p_\mu p_\nu $, 
and substituting (\ref{eq:pert.expansion})
in (\ref{eq:pert.eom}), yields
\bea
&& \begin{array}{ll}
\Gamma_T = - \xu  \,, & \qquad \Gamma_L + \gamma_{44} = 3 \xu \,, \cr
\curlh_{\mu\nu}'' - p^2 \curlh_{\mu\nu} = 0 \,, & \qquad
\gamma_{44}'' +  p^2 \gamma_{44}   =  4 \xu'' - p^2 \xu \,, \cr
\end{array}  \label{eq:pert.eqs} 
\eea and 
\beq
H_0 \chibf = -p^2 \chibf  \,;
\qquad H_0 = - \frac{ d^2}{d y^2} + S U \,,
\label{eq:chi.0}
\eeq
where the first equation in (\ref{eq:pert.eqs}) 
enforces the second condition in (\ref{eq:g.conditions}),
$S$ is defined in (\ref{eq:def.of.S}), and
\beq
 \xu = \frac{4 \bar\xi}{3|\aa|} \chi_1  \,,\qquad
U_{rs}(y) = - \inv{\sqrt{|m_r m_s|}}
\deriva{^2V}{\phi_r\, \partial\phi_s}{\hbox{\scriptsize$\phibf$}=\Phibf}
 \,.
\eeq
Destabilizing modes correspond to periodic (in $y$)
solutions to these
equations when $ p^2 > 0 $; in this case
(\ref{eq:pert.eqs}) require
$ \curlh_{\mu\nu} =0$, as  otherwise the solutions
would not be periodic. Note also that
the modes $ \chibf = \curlh_{\mu\nu} = \Gamma_T =0,
~\Gamma_L =  - \gamma_{44} $, with $ \gamma_{44}$ obeying the
homogeneous equation $ \gamma_{44}'' + p^2 \gamma_{44} =0$,
are apparently destabilizing, but these modes are coordinate
artifacts that can be eliminated by a transformation
of the form (\ref{eq:new.x1}).

Since the homogeneous solution to the $ \gamma_{44} $
equation in (\ref{eq:pert.eqs}) can be eliminated using 
(\ref{eq:new.x1}), we can assume that
$ \gamma_{44} $ is determined by $\xu \propto \aa\cdot\chibf$.
The stability of the background configuration then depends
only on whether (\ref{eq:chi.0}), 
the  equation satisfied by $ \chibf $,
has periodic solutions only for $ p^2 < 0 $. The
analysis of this equation is complicated by the fact that 
$SU$ is not Hermitian (for the usual definition of the
inner product), so that its similarity to the Schr\"odinger
equation is not very useful in this case 
\footnote{ $H_0$ is
Hermitian under the inner product 
 $ \vevof{\chibf_1 | \chibf_2}_S = \oint dy \, \chibf_1^\dagger S \chibf_2 $,
but, thought this
can be used to show that 
$ p^2 $ must be real, it does not
provide information about its sign since this inner product is not 
positive definite,
and the  arguments based on
the Sturm-Liouville theory~\cite{bg,MF}
are not applicable.}. Instead I argue as follows

The general solution to the equation $ m \Phibf'' = \FF $ in 
(\ref{eq:back.0}) depends on $2N$  constants
({\it e.g.} the values of $ \Phibf$ and
$ \Phibf' $ at $ y =0 $). The periodicity
requirement (\ref{eq:per.con}) introduces  $N$ restrictions
and one new parameter, since $L$ is not fixed {\it a priori}.
Finally, (\ref{eq:sum.rule}) imposes one additional constraint.
As a result the general periodic solution will depend on $N$ 
arbitrary parameters (including $L$); since the equation of motion does not
depends explicitly on $y$, one of these parameters can be taken as some
initial value $ y_0 $. The general periodic solution can
then be written
\beq
\Phibf = \Phibf(y-y_0; \cc ; L) \,, \quad
\cc = (c_1, \ldots , c_{N-2}) \,.
\eeq

Let now
\beq
\chibf\up0 = \partial_{y_0} \Phibf, \quad
\chibf\up{N-1} = \inv\epsilon \delta'_L \Phibf, \quad
\chibf\up i = \partial_{c_i} \Phibf\,, ~~ i=1, \ldots, N-2  \,;
\label{eq:zero.modes}
\eeq
where $ \delta'_L $ is defined below
(\ref{eq:delta.prime}). Then, by
taking the appropriate derivatives of the
equation of motion (\ref{eq:back.0}), it is easy
to see that
\beq
H_0 \chibf\up i =0 \,, ~~i=0, \ldots N-1 \,.
\eeq
Since the
constants $ y_0,~L$ and \cc\ are independent, the
modes $ \chibf\up i,~ i=0, \ldots, N-1 $ will be linearly 
independent (and periodic). 

The equation $ H_0 \chibf =0 $
is a linear, second-order differential equation for the
component functions $ \chi_r $, so the general solution
will contain $2N$ arbitrary constants.
The periodicity condition, however, imposes $N$ constraints
(note that now $L$ is fixed by the background solution),
so that the general periodic solution
will contain only $N$ independent constants. Then, since the equation
is linear, any periodic
solution to (\ref{eq:chi.0})
with $ p^2 =0 $ can be written as a linear combination of
the $ \chibf\up i $.

Consider next those background solutions that remain
near an extremum $ \Phibf_{\rm extr}$  of the potential $V$
(the presence
of an extremum is a necessary condition for the existence 
of periodic solutions (\ref{eq:per.con});
when $ m_1 < 0 $, as for the cases
of interest, this extremum is a saddle point).
For these cases the relevant equations are, approximately,
\beq
\tilde \Phibf'' = S \bar U \tilde \Phibf  \,, \qquad
- \chibf'' + S \bar U \chibf = - p^2 \chibf  \,;
\label{eq:small.ampl}
\eeq
where
\beq
 \tP_r = |m_r|^{1/2} \left(\Phibf
 - \Phibf_{\rm extr} \right)_r \,; \qquad 
\bar U_{r s} = - | m_r m_s|^{-1/2} 
\deriva{^2 V}{\phi_r \phi_s}{\hbox{\scriptsize$\phibf$} =\Phibf_{\rm extr} } \,.
\eeq
Now, for there to be periodic solutions
$ \tPbf$, $S \bar U $
must have at least one negative eigenvalue
$ - \omega^2$ ($\omega$ real); denoting
by $\vv$ the corresponding eigenvector,
it follows that the $y$-independent choice $ \chibf = 
\vv $ is a solution to the $\chibf$
equation with $ p^2 = \omega^2 > 0 $, which is then
a destabilizing mode. It follows that all small
amplitude solutions are unstable.

Now consider a small amplitude solution 
$ \Phi( y-y_0' ; \cc'; L') $, another arbitrary
solution $ \Phi( y-y_0'' ; \cc''; L'') $,
and  a smooth path in parameter space
$ \{ y_0(s) , \cc(s), L(s) \} $, $ 0 \le s \le 1 $, such that
$ \{ y_0(0) , \cc(0), L(0) \} = 
\{ y_0' , \cc', L' \} $ and
$ \{ y_0(1) , \cc(1), L(1) \} = 
\{ y_0'' , \cc'', L'' \} $. Each value of $s$
defines a periodic solution $ \Phibf$
with period $L(s)$,
form which a corresponding $U$ is constructed.
As a result all the modes $ \chibf $ will
depend smoothly on $s$, and so will the 
eigenvalues $ -p^2 $; in particular, the
$ p^2 =0 $ subspace is $N$ dimensional 
 for all $s$. For $ s =0 $ we know there is
at least one mode with $ p^2 > 0 $, but then this
mode must remain destabilizing for all $s$,
for if it were to change from destabilizing to stable,
there would be a value $ \bar s $ at which its 
eigenvalue $ p^2 $ would vanish, so that for $ s = \bar s $
the $ p^2 =0 $ subspace would have dimension $ N+1 $,
which we saw above is impossible. It follows that the
corresponding
solution $ \Phi( y-y_0'' ; \cc''; L'') $ is also unstable, and
since the parameters $ \{y_0'',\cc'',L''\}$ are arbitrary, it follows
that all background solutions in the pure scalar-gravity model
are unstable.

There is a subtlety in the above argument. One can easily imagine
potentials with two or more extrema, each with different number 
of destabilizing modes, yet the above argument seems to indicate that
the number of destabilizing modes cannot change. This apparent
contradiction is resolved by noting that there is no periodic
solution  that can interpolate between the corresponding
small-amplitude solutions. Imagine a potential with 2 extrema
and small-amplitude solution near each, denoted 
by $ \Phibf\up1(y) $ and $ \Phibf\up2(y) $,
and for which the corresponding $ \chibf $ equations 
in (\ref{eq:small.ampl}) have
different numbers of destabilizing modes. Then for any periodic
function $ \Phibf(y,s)$, $ 0 \le s \le 1 $, of period $L(s) $,
that solves (\ref{eq:back.0}) for each $s$, and
such that $ \Phibf\up1 = \Phibf|_{s=0},~ \Phibf\up2 = \Phibf|_{s=1} $,
there will be an intermediate
value $ 0< s_\infty<1 $ such that $ L(s) \to \infty  $ as $ s\to s_\infty$. 
Thus, to every extremum one can associated a ``region
of influence'' determined by all solutions that can be reached
by an interpolation with $L(s)$ finite for all $s$; the above argument
shows that all such solutions are unstable. I will assume that 
all periodic solutions can be characterized in this way, that is,
that any periodic solution can be ``deformed'' into a small
amplitude solution near an extremum of $V$ while keeping the period finite.
These results extend the arguments of \cite{bg,persol} to the
class of models considered here.

Similar considerations can be followed in case the potential is
not quadratic near the extremum.

\section{Antisymmetric tensor stabilization mechanism}
\label{sec:atf.stab}

The instability of the solutions to (\ref{eq:back.0}) is reminiscent
of the well-known instability of soliton-like solutions in more than $1+1$
dimensions \cite{sol}. In the soliton case stable solutions are obtained
by introducing gauge fields, here
I will pursue a different approach based on the introduction of an
antisymmetric tensor field $A$ that can propagate in the bulk, and which has the 
following interactions:
\bea
\lcal_A &=& - \inv{12} g^{ij} g^{kl} g^{mn} A_{ikm} A_{jln}
- \inv4 \kappa^2 g^{ij} g^{kl} A_{ik} A_{jl}
+ \half \kappa g^{ij} g^{kl} J_{ik} A_{jl}\,; \cr
J_{ik} &=& \sum_{r,s} | m_r m_s |^{1/2} \lambda_{rs}  \phi_{r,i} \phi_{s,k} \,,
\label{eq:l.A}
\eea
where $ A_{ijk} = 5 A_{[ij,k]} $ and $ \lambda_{rs} + \lambda_{sr} =0$
(the couplings $\lambda $ could depend on $ \phi$ also, but such a case
will not be considered here). When considering the quantum aspects of
this model it is convenient to rewrite $ \lcal_A $ using an auxiliary
St\"uckelberg-like vector field (see, for example, \cite{Kuzmin:2002gn});
the expression (\ref{eq:l.A}) then corresponds to the ``unitary'' gauge
where this auxiliary field vanishes.
The presence of higher-derivative terms, while innocuous classically,
can have dire quantum effects; still, since these theories have
an intrinsic UV cutoff, these are avoided
with appropriate constraints on the scales related to the couplings
$ \lambda $ (see sect. \ref{sec:comments}).
Antisymmetric tensors have been considered previously in the
literature in RS-like models with or without fundamental
scalars~\cite{AS-RS}, in non-periodic thick-brane
models with weak scalar-tensor couplings~\cite{Alencar:2010hs},
and in exotic-Lagrangian models addressing the self-tuning 
of the cosmological constant in RS-like models~\cite{AS-CC}.
The type of models described by (\ref{eq:l.A}) has apparently not
been discussed previously within the present context.

I will assume that $ \kappa = O ( \varphi^{2/3} ) $ so that $A \sim 1/\varphi $;
then
\beq
\lcal_A = - \inv 4 (\kappa A - J )_{ij} ( \kappa A - J)^{ij} + \inv4 J_{ij} J^{ij}
+ O (1/\varphi)\,,
\eeq
where indices are raised and lowered using the flat space metric $ \eta_{ij}$.
In this case the field $A$ can be integrated out and one can work instead
with the effective Lagrangian
\beq
\lcal_{\rm eff} = \inv4 J_{ij} J^{ij}\,,
\label{eq:leff}
\eeq
up to corrections of order $ 1/\varphi $.

The addition of $ \lcal_{\rm eff} $ does not change the background equations
(to lowest order) so that the results of sections \ref{sec:bs}-\ref{sec:rs}
are not modified; in terms of the antisymmetric
tensor field this implies that it vanishes in the background configuration.

\subsection{Stability with antisymmetric tensors}

The perturbation equation {\it is} modified by the addition
of (\ref{eq:leff}). Instead of (\ref{eq:chi.0}) one gets
\beq
H_0 \chibf = - p^2 \left[ \chibf - S \BB (\BB \cdot \chibf) \right] \,,
\quad B_r = \sum_s |m_s|^{1/2} \lambda_{rs} \Phi_s'\,.
\label{eq:pert.A}
\eeq
As before the perturbative stability of the background configuration 
$ \Phibf $ is guaranteed if this equation has no solutions for
$ p^2>0 $. 

Given the
form of (\ref{eq:pert.A}), it is clear that the modes
$ \chibf\up i $ defined in (\ref{eq:zero.modes}) again provide
$N$ solutions corresponding to $ p^2 =0 $. The argument 
of section \ref{sec:stability} then implies
that the background configurations will be stable provided
the small-amplitude
background configurations near an extremum are stable; this
can be realized when $ \BB \not =0 $.

For background configurations that remain near an extremum of 
the potential $V$ we  still have  $ \tPbf'' = S \bar U \tilde \Phibf $ 
as in (\ref{eq:small.ampl}), but the equation for $ \chibf $ becomes
\beq
- \chibf'' + S \bar U \chibf  = - p^2 \chibf + 
p^2 S \bar\BB ( \bar\BB \cdot \chibf) \,,
\quad \bar B_r = \sum_s \lambda_{rs} \tP_s' \,.
\label{eq:chi.A}
\eeq
In order to have periodic solutions $ \tPbf $,
$ S \bar U $ must have one or more negative eigenvalues
$ - \omega_a^2 $, with $ \vv_a $
the corresponding eigenvectors; 
however, the constant modes $ \chibf = \vv_a $ are not
solutions of (\ref{eq:chi.A}) when $\BB \not =0 $.

I will now consider the 
case where the potential takes the form (\ref{eq:simple.v}), with
$u_r $ of the shape given in Fig. \ref{fig:i} for all $r$.
Then each $ \tP_r'$ will be approximately a square wave, so that I can
approximate $ (\tP_r')^2 \to a_r = $ constant, and 
the sum rule (\ref{eq:sum.rule}) becomes
\beq
a_1 = \sum_{r>1} a_r \,.
\eeq
In addition,
$ (S\bar U )_{r s} = \delta_{rs} \vcal_r $ is diagonal, and 
each $ \vcal_r $ can be approximated by a delta-function comb,
with negative magnitude and half the period of $ \tP_r $.

I will now make the simplifying assumption that the periods
of $ \tP_r $ for $ r > 1 $ are much smaller than that
of $ \tP_1 $; for example, if $ \ell $ denotes the period of $\tP_1 $,
that of $\tP_r,~ r>1 $ can be chosen to be $ \ell/n_r $ with
$ n_r $ a large prime number; the period
$L$ in (\ref{eq:per.con}) then equals $ \left(\prod n_r \right) \ell $.
In this case $ S\bar U $ and
$ \bar\BB \otimes \bar\BB $ in (\ref{eq:chi.A}) will have
entries that are slowly varying and others that vary very rapidly,
and this equation can be treated using the procedure described
in~\cite{h-w}: up to corrections of order $ 1/n_r $, the solution
can be obtained by considering only the slowly varying terms in 
(\ref{eq:chi.A}):
\beq
- \chi_r''  - \delta_{r,1} \vcal_1 \chi_1 = - p^2 \chi_r + p^2 S_r 
\sum_s\bcal_{rs} \chi_s
\label{eq:av.eq}
\eeq
(no sum over $r$) where, as before, $ S_1 = -1 ,~ S_{r>1} = + 1 $,  and
\beq
\bcal_{rs} = \sum_{u>1}\left( \lambda_{r1} \lambda_{s1} 
+ \lambda_{ru} \lambda_{su}
\right) a_u \,.
\eeq
The term containing $ \bcal $ in (\ref{eq:av.eq}) is generated
by averaging the rapidly varying terms in (\ref{eq:chi.A})
over periods $ \ll \ell $.

\begin{figure}
\begin{center}
\includegraphics[width=6in]{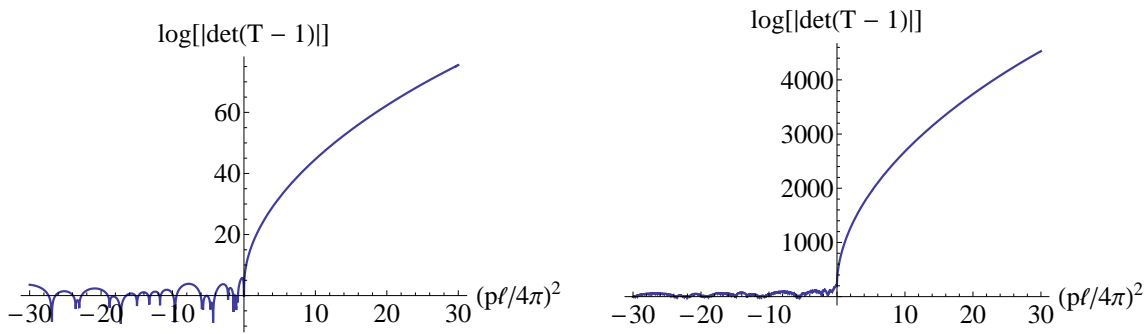}
\end{center}
\caption{Value of $\log \left| \hbox{det}(T - \mati) \right| $ 
when $ \ncal = 1 $ (left) and $ \ncal = 60 $ (right)
In both cases
$ \lambda_{12} = - 0.2,~ \lambda_{13} = - 0.4,~ \lambda_{23} = 0.3,~
a_2 = 0.05,~ a_3 = 0.7 $, and the amplitude of $ \vcal_1 $ equals
$28 \pi /\ell $. Periodic solutions occur for $ p^2 \le 0 $
and do not correspond to destabilizing modes.}
\label{fig:ii}
\end{figure}

\begin{figure}
\begin{center}
\includegraphics[width=3in]{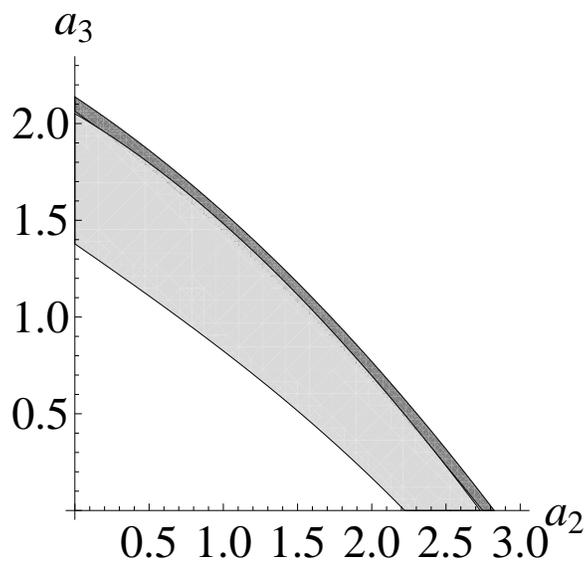}
\end{center}
\caption{Values of $a_{2,3}$ for which (\ref{eq:av.eq})
has no destabilizing modes. Light gray: $ \wp < 30 $ and 
the amplitude of $ \vcal_1 $ equals $ 28 \pi /\ell $;
dark gray: $ \wp < 20 $ and 
the amplitude of $ \vcal_1 $ equals $ 4 \pi /\ell $.
In both cases
$ \lambda_{12} = - 0.2,~ \lambda_{13} = - 0.4,~ \lambda_{23} = 0.3 $}
\label{fig:iii}
\end{figure}

The equation for $ \chi_ r $ can now by analyzed using 
standard techniques (see e.g. \cite{grim}): denoting
$ \XX = ( \chibf , \chibf')^t $ there will be a non-singular matrix
$T$ such that $ \XX(y + L) = T \XX(y) $, and the system
will support destabilizing modes if, for some {\it positive} $ p^2 $,
$T$ has an eigenvalue equal to one (which corresponds to 
$ \chibf $ being periodic). In determining the conditions under which
this occurs one is interested only in modes for which $ |p^2| < \Lambda^2 $,
or, equivalently, $ [ p \ell/(4\pi)]^2 \le [\Lambda L/ (4 \pi \ncal)]^2 
= \wp $, where $ \ncal = \prod n_r $ and 
$ \ell /n_r $ is the period of $ \tP_r $, defined above; 
with appropriate choice of these parameters one can insure
$\wp = \Or( 10 ) $. In this case there are ranges of 
parameters where not destabilizing modes are present;  
Fig. \ref{fig:ii} contains an example for fixed $ \lambda_{rs},~a_r $ and $ \vcal_1 $
where the matrix $T \not = \mati$ for $p^2 > 0 $, while 
Fig. \ref{fig:iii} gives, for fixed $ \lambda_{r,s} $, the values of $ a_{2,3} $ 
where no destabilizing modes occur (within the allowed $p^2 $ range).
Background configurations for this type of models are then stable in such cases.

Stable solutions occur when the number of scalar fields
is $ \ge 3 $;  I found no stable configurations in more economical 
models.

\section{Comments}
\label{sec:comments}

The main purpose of the present paper was to provide mechanisms that
can overcome the periodicity and 
stability problems of brane-like configurations
produced by scalar fields. The model defined by (\ref{eq:l.phi}) and
(\ref{eq:l.A})
meets these requirements by introducing a coupling of
the scalars to the Ricci scalar and to 
an antisymmetric tensor field that can propagate
in the bulk. For appropriate choices of scalar potential the
background configurations mimic that of the RS model but
require the presence of $ \ge3$ scalar fields with modes
of very different though commensurate periods 
(an investigation of whether this hierarchy can be maintained
naturally lies beyond the scope of this paper).

As mentioned in section \ref{sec:intro} these models 
contain a UV cutoff scale $ \Lambda $; 
when $\bBB$ is compact they also
contain another high-energy scale, the compactification
radius $L$. An immediate concern is whether one can naturally assume $
\Lambda \gg 1/L$, for otherwise the model cannot accurately describe
the dynamics associated with the compact directions.  This question
can be investigated using an extension of naive dimensional analysis
\cite{NDA}. For example, a straightforward estimate for 5-dimensional
gauge theories gives
$ \Lambda \sim 24\pi^3/g_5^2 $, where $g_5$ is the 5-dimensional gauge
coupling constant (which is dimensional). The 4-dimensional gauge
coupling is $g \sim g_5/\sqrt{L}$, whence $ L \Lambda \sim 24 \pi^3/g^2$, 
which indicates that such models are reliable at
energies above the compactification
scale provided $  6 \pi^2  \gg g^2/(4\pi)$. 
For compact $\bBB$ the scale $\Lambda $ also provides a
cutoff for the order of the KK excitations that need to be included in
the theory; the effects of higher order modes are
absorbed into a renormalization of the various operator coefficients.

When the antisymmetric tensor field is introduced,
additional constraints appear, such as, for example,
those derived from the unitarity constraint
in $ \phi \phi \to \phi \phi $ scattering, which
demands (at tree-level) that the scale
associated with $ \lambda $ be below $ \sim \Lambda/(2\pi) $;
this restriction also insures that the  effective Lagrangian
(\ref{eq:leff}) does not generate undesirable poles in the scalar  propagators.

In the above models there are no exponentially destabilizing
modes below scale $ \Lambda $. There {\em are}, however, zero modes
 obtained by taking derivatives
with respect to the parameters of the background configuration
(see Eq. \ref{eq:zero.modes}).
Perturbations along these zero modes then correspond
to small deformations of the background configuration parameters,
under which its period and orbits 
suffer small changes, but and do not result in an instability.

Phenomenologically viable models constructed along these lines must
include, in addition, the possibility of orbifolding. In this case
it is most convenient to assume that the potential allows solutions 
with definite parity under $ y \leftrightarrow - y $. This can be
implemented without additional complications.

It is worth noting that the present models cannot be stabilized
by the 1-loop effective potential \cite{rad}. This is because
in the present case the tree-level potential $U$ supports at least
one destabilizing mode (whose amplitude increases exponentially
with time), not merely a neutral one. Loop corrections are of course present, 
but they are subdominant.

\appendix

\section{Alternative interpretation of the scalar equations of motion}

\label{sec:app.c}

When $ \bar\xi > 3/16 $, the 
presence of a negative sign in $S$ 
(cf. eq. \ref{eq:def.of.S}) suggests an alternative
description of the equations for $ \Phibf$ as a geodesic
equation, and that for $ \chibf$ as a geodesic deviation equation.

To see this consider an $ N+1$-dimensional space with coordinates
$ \Phi_1, \ldots, \Phi_N, \theta $ and metric
\beq
\Gamma_{R S} = {\rm diag}\left( m_1,m_2,\ldots,m_N,1/V(\Phibf) \right) \,.
\label{eq:met}
\eeq
I will denote by $r,s,$ etc. the indices corresponding 
to the first $N$ coordinates, then $\Gamma_{rs} = m_r \delta_{rs} $. The
geodesic equations associated with this metric are
\beq
\frac{d^2 \Phi_r}{d p^2} = - \inv{2 m_r V^2}\deriva V{\Phi_r}{}
\left( \frac{d\theta}{d p} \right)^2 \,;
~
\frac{d^2 \theta}{d p^2} = \inv V \sum_r \deriva V{\Phi_r}{}
\left( \frac{d\Phi_r}{d p} \right)
\left( \frac{d\theta}{d p} \right)
; 
\eeq
where $p$ is an affine parameter.

The second equation can be immediately integrated:
$ d\theta/d p = c V$ where $c$ is a constant; substituting
into the equation for the $ \Phi_r $ gives
\beq
\frac 2{c^2} \frac{d^2 \Phi_r}{d p^2} = -\deriva V{\Phi_r}{} \,,
\eeq
that reduces to the second-order
equation for $ \Phibf$ in (\ref{eq:back.0})
provided $y$ is identified with $ c p/\sqrt{2} $. 

The zero-energy condition in (\ref{eq:back.0}) corresponds
to the requirement that this be a null geodesic:
\beq
\sum_{rs} \frac{d\Phi_r}{d p}  \frac{d\Phi_s}{d p} \Gamma_{rs}
+ \inv V \left( \frac{d\theta}{d p} \right)^2 =
c^2 \left\{
\half \Phibf' m \Phibf' + V(\Phibf) \right\} = 0 \,,
\eeq
where a prime denotes a $y$ derivative.
It is also clear that the equation for $ \chibf$ is
equivalent to the geodesic deviation equation 
associated with the metric (\ref{eq:met}).

\section{Background solutions for models
with a single scalar field.}

\label{sec:app.a}

When there is a single scalar field the background solution
takes the form 
\bea
\tilde\phi &=& \Phi + \Theta/\varphi + \cdots \cr
g_{\mu \nu} &=& \left( 1 - 2 \sigma/\varphi -2 \tau/\varphi^2 + \cdots
\right) \eta_{\mu\nu}
\eea
with $ g_{44}=1,~ g_{4\mu} =0 $. 

For this simple case the field equation is redundant; while the 
Einstein equations give
\bea
 O(\varphi): && 2 \zeta \Phi'' - 3 \sigma'' =0 \,,\cr
 O(1): &&
V(\Phi) - \half m \Phi'{}^2  
- \frac{2\xi}{3\zeta} \left( 2\zeta \Phi' - 3 \sigma' \right)^2 =0 \,,\cr
&& 2  \Theta'' - \frac3\zeta \tau''
- 6  \Phi \sigma'' 
+ 2  \Phi' \sigma'
- \inv\xi \Phi'{}^2 + \left( \Phi^2 \right)'' =0 \,,\cr
 O(1/\varphi): &&
 4 \xi \Phi \left( 2\Phi - 3 \sigma \right)' \sigma'
- \frac{12\xi}\zeta \sigma' \tau'
+ \left( 8 \xi \tau - \Theta \right)' \Phi' 
+ 8 \xi \sigma' \Theta' + \Theta
\deriva V\phi{\Phi} =0 \cr &&
\label{eq:spp}
\eea
(I omitted an additional equation to order $ 1/\varphi$ that determines the
correction of order $ 1/\varphi^3$ to the metric), where
\beq
\bar\xi = \frac{\xi^2}{\xi - 4\mpl^3/\varphi^2}\,, \qquad
\zeta = \bar\xi/\xi \,, \qquad m = 1 - \frac{16}3 \bar\xi  \,.
\eeq

The $ O(\varphi)$ equation together with the periodicity requirement
imply $ 2 \zeta \Phi - 3 \sigma= $ constant, whence $ \Phi $ must satisfy
\beq
\half m \Phi'{}^2 + V\left(\Phi \right) =0 \,.
\label{eq:zero.energy}
\eeq
It now proves convenient to write
\bea
\Theta &=& \Phi' \, A(y) \,, \cr
\tau &=& \inv{2\xi} f\left(\Phi\right) 
+ \frac{2\zeta}3 \left( \frac{1-\zeta}m - \half
\right) \Phi^2  + B(y) \,,
\eea
Substituting into (\ref{eq:spp}) and using
(\ref{eq:zero.energy}) one finds
\bea
\sigma &=& \frac{2 \zeta}3  \Phi \,; \qquad
A'=   \frac{(m-1)(1-\zeta)}m \Phi \,;\cr
B'' &=& \frac{2 \zeta}3 A \Phi''' \,; \qquad
f'' + \half  \frac{V'}V f'  + \frac{2 \zeta}3
\left[ m
+ \frac{32\xi^2}3 \frac{\zeta(1-\zeta)}m \right]  =0 \,;
\eea
where a prime denotes a derivative with respect to the argument.
These equations can be solved by quadratures; in particular,
\beq
f(\phi) = - 
\frac{2 \zeta}{9m}
\left[3m^2 
+ 32 \bar \xi (\xi - \bar\xi)  \right]
\int_0^\phi d\lambda  \int_0^\lambda d\gamma \sqrt{
\frac{ V(\gamma)}{V(\lambda)}} \,.
\eeq

Higher orders can be dealt with similarly.
For more than one field the higher-order
corrections to (\ref{eq:back.0}) cannot, in general,
be cast into such comparatively simple expressions.

\section*{Acknowledgments}
The author wishes to thank J.L. Padilla who was involved in the early states of this
project; and B. Grzadkowski for illuminating comments. This work
was supported in part by the U. S. Department of Energy under Grant No. DEFG03- 94ER40837


\section*{References}
\begin{thebibliography}{99}


\bibitem{rs}
  Randall L and Sundrum R 1999 {\it Phys. Rev. Lett.} {\bf 83} 3370  ({\it Preprint} arXiv:hep-ph/9905221); 
  Randall L and Sundrum R 1999 {\it Phys. Rev. Lett.} {\bf 83} 4690 ({\it Preprint} arXiv:hep-th/9906064).

\bibitem{phen}
  Huber S~J~ and Shafi Q 2001 {\it  Phys. Lett.} B {\bf 498} 256 ({\it Preprint} arXiv:hep-ph/0010195).
  Agashe K {\it et al.} 2007 {\it  Phys. Rev.} D {\bf 76} 115015 ({\it Preprint} arXiv:0709.0007 [hep-ph]).
  Bouchart C and Moreau G 2009 {\it Nucl. Phys.} B {\bf 810} 66 ({\it Preprint} arXiv:0807.4461 [hep-ph]).
  Carena M, Medina A D, Shah N R and Wagner C E M 2009 {\it Phys. Rev.} D {\bf 79} 096010 ({\it Preprint} arXiv:0901.0609 [hep-ph]).

\bibitem{d.6}
  Nelson A E 2001 {\it Phys. Rev.} D {\bf 63} 087503 ({\it Preprint} arXiv:hep-th/9909001).
  Cohen A G and Kaplan D B 1999 {\it Phys. Lett.}  B {\bf 470} 52 ({\it Preprint} arXiv:hep-th/9910132).
  Hawking S W, Hertog T and Reall H S 2000 {\it  Phys. Rev.} D {\bf 62} 043501 ({\it Preprint} arXiv:hep-th/0003052).
  Gogberashvili M and Midodashvili P 2001 {\it Phys. Lett.}  B {\bf 515} 447 ({\it Preprint} arXiv:hep-ph/0005298).
  Duff M J, Liu J T and Stelle K S 2001 {\it J. Math. Phys.} {\bf 42} 3027 ({\it Preprint} arXiv:hep-th/0007120).
  Collins H and Holdom B 2001 {\it Phys. Rev.} D {\bf 64} 064003 ({\it Preprint} arXiv:hep-ph/0103103).
  Gibbons G W and Hull C M 2001 arXiv:hep-th/0111072.
  Bergshoeff E, Gran U and Roest D 2002 {\it Class. Quant. Grav.} {\bf 19} 4207 ({\it Preprint} arXiv:hep-th/0203202).
  Nishino H and Rajpoot S 2002 {\it Phys. Lett.} B {\bf 546} 261 ({\it Preprint} arXiv:hep-th/0207246).
  Carroll S M and Guica M M 2003 arXiv:hep-th/0302067.
  Guendelman E I 2004 {\it Phys. Lett.  B} {\bf 580} 87 ({\it Preprint} arXiv:gr-qc/0303048).
  Cline J M, DescheneauJ , Giovannini M and Vinet J 2003 {\it JHEP} {\bf 0306} 048 ({\it Preprint} arXiv:hep-th/0304147).
  Bao R and Lykken J D 2005 {\it Phys. Rev. Lett.} {\bf 95} 261601 ({\it Preprint} arXiv:hep-ph/0509137).
  Dzhunushaliev V, Folomeev V, Myrzakulov K and Myrzakulov R 2009 {\it Gen. Rel. Grav.} {\bf 41} 131 ({\it Preprint} arXiv:0705.4014 [gr-qc]).

\bibitem{NDA}
  Manohar A and Georgi H 1984 {\it Nucl. Phys.} B {\bf 234} 189.
  Georgi H 1993 {\it Phys. Lett.} B {\bf 298} 187 ({\it Preprint} arXiv:hep-ph/9207278).
  Chacko Z, Luty M A and Ponton E 2000 {\it JHEP} {\bf 0007} 036 ({\it Preprint} arXiv:hep-ph/9909248).
  Grzadkowski B and Wudka J, {\it Phys. Rev.} D {\bf 77} 096004 ({\it Preprint} arXiv:0705.4307 [hep-ph]).

\bibitem{add}
  Rubakov V A and Shaposhnikov M E 1983 {\it Phys. Lett.} B {\bf 125} 136.
  Arkani-Hamed N, Dimopoulos S and Dvali G R {\it Phys. Lett.} B {\bf 429} 263 ({\it Preprint} arXiv:hep-ph/9803315).
  Antoniadis I, Arkani-Hamed N, Dimopoulos S and Dvali G R {\it Phys. Lett.} B {\bf 436} 257 ({\it Preprint} arXiv:hep-ph/9804398).
  Shiu G and Tye S H H {\it Phys. Rev.} D {\bf 58} 106007 ({\it Preprint} arXiv:hep-th/9805157).
  Sundrum R 1999 {\it Phys. Rev.} D {\bf 59} 085009 ({\it Preprint} arXiv:hep-ph/9805471).
  Kakushadze Z and Tye S H H 1998 {\it Nucl. Phys.} B {\bf 548} 180 ({\it Preprint} arXiv:hep-th/9809147).
  Nussinov S and Shrock R 1999 {\it Phys. Rev.} D {\bf 59} 105002 ({\it Preprint} arXiv:hep-ph/9811323).
  Donini A and Rigolin S 1999 {\it Nucl. Phys.}  B {\bf 550} 59 ({\it Preprint} arXiv:hep-ph/9901443).
  For a recent review see: Shifman M 2009 {\it Preprint} arXiv:0907.3074 [hep-ph].

\bibitem{rv}
  Davies R, George D P and Volkas R R 2008 {\it Phys. Rev.} D {\bf 77} 124038 ({\it Preprint} arXiv:0705.1584 [hep-ph]).

\bibitem{rad.stab}
  Csaki C, Graesser M, Randall L and Terning J 2000 {\it Phys. Rev.} D {\bf 62} 045015 ({\it Preprint} arXiv:hep-ph/9911406).

\bibitem{gw}
  Goldberger W D and Wise M B 1999 {\it Phys. Rev. Lett.} {\bf 83} 4922 ({\it Preprint} arXiv:hep-ph/9907447).

\bibitem{BSMP}
  DeWolfe O, Freedman D Z, Gubser S S and Karch A 2000 {\it Phys. Rev.}  D {\bf 62} 046008 ({\it Preprint} arXiv:hep-th/9909134).
  Goldberger W D and Wise M B {\it Phys. Lett.} B {\bf 475} 275 ({\it Preprint} arXiv:hep-ph/9911457).
  Gremm M 2000 {\it Phys. Lett.} B {\bf 478} 434 ({\it Preprint} arXiv:hep-th/9912060); {\it Phys. Rev.}  D {\bf 62} 044017 ({\it Preprint} arXiv:hep-th/0002040).

\bibitem{BSNP}
  Goldberger W D and Wise M B 2000 {\it Phys. Lett.} B {\bf 475} 275 ({\it Preprint} arXiv:hep-ph/9911457).
  Grzadkowski B and Gunion J F 2003 {\it Phys. Rev.} D {\bf 68} 055002 ({\it Preprint} arXiv:hep-ph/0304241).
  Farakos K and Pasipoularides P 2006 {\it Phys. Rev.} D {\bf 73} 084012 ({\it Preprint} arXiv:hep-th/0602200).
  Bogdanos C, Dimitriadis A and Tamvakis K 2006 {\it Phys. Rev.  D} {\bf 74} 045003 ({\it Preprint} arXiv:hep-th/0604182).
  Farakos K, Koutsoumbas G and Pasipoularides P 2007 {\it Phys. Rev.  D} {\bf 76} 064025 ({\it Preprint} arXiv:0705.2364 [hep-th]).

\bibitem{BSBP}
  Mikhailov A S, Mikhailov Yu S, Smolyakov M N and Volobuev I P 2008 {\it Preprint} arXiv:0812.2699 [hep-th].

\bibitem{NSMNN}
  Chacko Z and Nelson A E {\it Phys. Rev.} D {\bf 62} 085006 ({\it Preprint} arXiv:hep-th/9912186).
  Kehagias A and Tamvakis K 2001 {\it Phys. Lett.} B {\bf 504} 38 ({\it Preprint} arXiv:hep-th/0010112).
  Ringeval C, Peter P and Uzan J P 2002 {\it Phys. Rev.} D {\bf 65} 044016 ({\it Preprint} arXiv:hep-th/0109194).
  Guerrero R, Melfo A and Pantoja N 2002 {\it Phys. Rev.} D {\bf 65} 125010 ({\it Preprint} arXiv:gr-qc/0202011).
  Melfo A, Pantoja N and Skirzewski A 2003 {\it Phys. Rev.} D {\bf 67} 105003 ({\it Preprint} arXiv:gr-qc/0211081).
  Bronnikov K A and Meierovich B E 2003 {\it Grav. Cosmol.} {\bf 9} 313 ({\it Preprint} arXiv:gr-qc/0402030).
  Castillo-Felisola O, Melfo A, Pantoja N and Ramirez A 2004 {\it Phys. Rev.} D {\bf 70} 104029 ({\it Preprint} arXiv:hep-th/0404083).

\bibitem{NSMNS}
  DeWolfe O, Freedman D Z, Gubser S S and Karch A 2000 {\it Phys. Rev.} D {\bf 62} 046008 ({\it Preprint} arXiv:hep-th/9909134).
  DeWolfe O and Freedman D Z 2000 {\it Preprint} arXiv:hep-th/0002226.
  Kehagias A and Tamvakis K 2001 {\it Phys. Lett.} B {\bf 504} 38 ({\it Preprint} arXiv:hep-th/0010112).
  Kehagias A and Tamvakis K 2002 {\it Mod. Phys. Lett.} A {\bf 17} 1767 ({\it Preprint} arXiv:hep-th/0011006).
  Giovannini M 2001 {\it Phys. Rev.} D {\bf 64} 064023 ({\it Preprint} arXiv:hep-th/0106041).
  Giovannini M 2002 {\it Phys. Rev.} D {\bf 65} 064008 ({\it Preprint} arXiv:hep-th/0106131).
  Kobayashi S, Koyama K and Soda J 2002 {\it Phys. Rev.} D {\bf 65} 064014 ({\it Preprint} arXiv:hep-th/0107025).
  Ghoroku K and Yahiro M 2003 {\it Preprint}  arXiv:hep-th/0305150.
  Boos E E, Mikhailov Y S, Smolyakov M N and Volobuev I P 2005 {\it Nucl. Phys.} B {\bf 717} 19 ({\it Preprint} arXiv:hep-th/0412204).
  Afonso V I, Bazeia D and Losano L 2006 {\it Phys. Lett.} B {\bf 634} 526 ({\it Preprint} arXiv:hep-th/0601069).
  Andrianov A A and Vecchi L 2008 {\it Phys. Rev.} D {\bf 77} 044035 ({\it Preprint} arXiv:0711.1955 [hep-th]).

\bibitem{pot}
  Wang A 2002 {\it  Phys. Rev.} D {\bf 66} 024024 ({\it Preprint} arXiv:hep-th/0201051).

\bibitem{NMMNN}
  Dando G, Davidson A, George D P, Volkas R R and Wali K C 2005 {\it Phys. Rev.}  D {\bf 72} 045016 ({\it Preprint} arXiv:hep-ph/0507097).
  Dzhunushaliev V, Schmidt H J, Myrzakulov K and Myrzakulov R 2006 {\it Preprint} arXiv:gr-qc/0610100.
  Du M, Du X and Xie Y 2008 {\it Mod. Phys. Lett.} A {\bf 23} 3179.


\bibitem{NMMNS}
  Dzhunushaliev V, Folomeev V and Minamitsuji M 2009 {\it Phys. Rev.} D {\bf 79} 024001 ({\it Preprint} arXiv:0809.4076 [gr-qc]).


\bibitem{srules}
  Gibbons G W, Kallosh R and Linde A D 2001 {\it JHEP} {\bf 0101} 022 ({\it Preprint} arXiv:hep-th/0011225).
  Leblond F, Myers R C and Winters D J 2001 {\it JHEP} {\bf 0107} 031 ({\it Preprint} arXiv:hep-th/0106140).
  See also: Ellwanger U 2000 {\it Phys. Lett.} B {\bf 473} 233 ({\it Preprint} arXiv:hep-th/9909103).
  Forste S, Lalak Z, Lavignac S and Nilles H P 2000 {\it JHEP} {\bf 0009} 034 ({\it Preprint} arXiv:hep-th/0006139).

\bibitem{bg}
  Grzadkowski B and Toharia M 2004 {\it Nucl. Phys.} B {\bf 686} 165 ({\it Preprint} arXiv:hep-ph/0401108).
  Toharia M and Trodden M 2008 {\it Phys. Rev. Lett.} {\bf 100} 041602 ({\it Preprint} arXiv:0708.4005 [hep-ph]).
  Toharia M and Trodden M 2008 {\it Phys. Rev.} D {\bf 77} 025029 ({\it Preprint} arXiv:0708.4008 [hep-ph]).

\bibitem{persol}
  Bakas I and Sourdis C 2002 {\it Fortsch. Phys.} {\bf 50} 815 ({\it Preprint} arXiv:hep-th/0205007).

\bibitem{Jackiw:1977yn} Jackiw R 1977 {\it  Rev.\ Mod.\ Phys.}  {\bf 49} 681.


\bibitem{ll}
  Landau L D and Lifshitz E M 2005 {\it The classical theory of fields. Course of theoretical physics Vol. 2} (Butterworth-Heinemann, Oxford) p 260

\bibitem{sol}
  For a review see: Jackiw R 1977 {\it Rev. Mod. Phys.} {\bf 49} 681.
  For an early construction of kink configurations, see:  Dashen R F, Hasslacher B and Neveu A 1974 {\it Phys. Rev.} D {\bf 10} 4130.


\bibitem{MF}
  Morse P M and Feshbach H 1953 {\it Methods of theoretical physics} (McGraw-Hiill, New York) p 719.

\bibitem{dk}
  Dennery P and Krzywicki A 1967 {\it Mathematics for Physicists} (Harper and Row, New York) p 165.

\bibitem{rad}
  For a review see: Sundrum R 2005 In: {\it Physics in $ D \ge 4$. Proceedings of the Theoretical Advanced Study Institute in Elementary Particle Physics; TASI 2004}. 
  Boulder, CO, USA, 6 June-2 July 2004. Edited by J. Terning; C.E.M. Wagner and D. Zeppenfeld (World Scientific; Hackensack, NJ; 2006; eBook)  arXiv:hep-th/0508134.

\bibitem{Kuzmin:2002gn}
Kuzmin  S V and  McKeon D G C 2002 {\it  Can.\ J.\ Phys.}  {\bf 80} 767.

\bibitem{AS-RS}
  SenGupta S and Sur S 2004 {\it Europhys. Lett.} {\bf 65} 601 ({\it Preprint} arXiv:hep-th/0306048).
  Mukhopadhyaya B, Sen S, Sen S and SenGupta S 2004 {\it Phys. Rev.} D {\bf 70} 066009 ({\it Preprint} arXiv:hep-th/0403098).
  Das S, Dey A and SenGupta S 2006 {\it Class. Quant. Grav.}  {\bf 23} L67 ({\it Preprint} arXiv:hep-th/0511247).
  Das S, Dey A and SenGupta S 2008 {\it Europhys. Lett.} {\bf 83} 51002 ({\it Preprint} arXiv:0704.3119 [hep-th]).
  Mukhopadhyaya B, Sen S and SenGupta S 2007 {\it Phys. Rev.} D {\bf 76} 121501 ({\it Preprint} arXiv:0709.3428 [hep-th]).
  Tahim M O, Cruz W T and Almeida C A S 2009 {\it Phys. Rev.} D {\bf 79} 085022 ({\it Preprint} arXiv:0808.2199 [hep-th]).
  Mukhopadhyaya B, Sen S and SenGupta S 2009 {\it Phys. Rev.} D {\bf 79} 124029 ({\it Preprint} arXiv:0903.0722 [hep-th]).
  Das A and SenGupta S 2010 {\it Preprint}  arXiv:1010.2076 [hep-th].
  Alencar G, Landim R R, Tahim M O and Mendes K C 2010 {\it Preprint} arXiv:1009.1183 [hep-th].
  Alencar G, Landim R R, Tahim M O, Muniz C R and Costa Filho R N 2010 {\it Phys. Lett.} B {\bf 693} 503 ({\it Preprint} arXiv:1008.0678 [hep-th]).
  Alencar G, Landim R R, Tahim M O, Muniz C R and Costa Filho R N 2010 {\it Preprint}  arXiv:1005.1691 [hep-th].

\bibitem{Alencar:2010hs}
  Alencar G, Landim R R, TahimM O, Muniz C R and Costa Filho R N {\it Phys. Lett.} B {\bf 693} 503 ({\it Preprint} arXiv:1008.0678 [hep-th]).

\bibitem{AS-CC}
  Kim J E, Kyae B and Lee H M 2001 {\it Phys. Rev. Lett.} {\bf 86} 4223 ({\it Preprint} arXiv:hep-th/0011118); {\it Nucl. Phys.} B {\bf 613} 306 ({\it Preprint} arXiv:hep-th/0101027).

\bibitem{h-w}
  Huerta R and Wudka J 2001 {\it Phys. Rev.} A {\bf 63} 062104.

\bibitem{grim}
  R. Grimshaw 1993 {\it Nonlinear ordinary differential equations} (CRC Press, Boca Raton) p 47.




\end{thebibliography}
\end{document}